# Dynamical light control in longitudinally modulated segmented waveguide arrays

Yaroslav V. Kartashov[1] and Victor A. Vysloukh[2]

[1]ICFO-Institut de Ciencies Fotoniques, and Universitat Politecnica de Catalunya, Mediterranean Technology Park, 08860 Castelldefels (Barcelona), Spain

[2]Departamento de Fisica y Matematicas, Universidad de las Americas – Puebla, Santa Catarina Martir, 72820, Puebla, Mexico

We address light propagation in segmented waveguide arrays where the refractive index is longitudinally modulated with an out-of-phase modulation in adjacent waveguides, so that the coupling strength varies along propagation direction. Thus in resonant segments coupling may be inhibited hence light remains localized, while in detuned segments coupling results in complex switching scenarios that may be controlled by stacking several resonant and nonresonant segments. By tuning the modulation frequency and lengths of waveguide segments one may control the distribution of light among the output guides, including localizing all light in the selected output channel.

   A longitudinal refractive index modulation is exceptionally powerful tool to control the evolution of light in waveguiding systems ranging from simplest couplers to topologically sophisticated waveguide arrays. This is possible because longitudinal refractive index modulation affects propagation constants of the eigenmodes of the guiding system (for example, Bloch waves in periodic waveguide arrays) in a very specific fashion, making them equal for specific modulation parameters. In particular, a longitudinal bending of individual waveguides modifies the strength of their coupling, resulting in the dynamic localization of light. This effect was predicted in a simplest two-channel system [1,2] and later observed in truly periodic arrays [3-6]. Similarly, the coupling between waveguides can be strongly altered by a shallow out-of-phase modulation of refractive index in neighboring guides [7-9], as it was observed in simplest arrays [10,11] and proposed in more sophisticated guiding structures [12,13]. The concept of control of localization of laser radiation by longitudinal refractive index modulations was successfully extended not only to two-dimensional geometries [14-18], but also to fully three-dimensional settings [19,20]. However, in most of the settings mentioned above only the simplest harmonic refractive index variations were considered although the facts that dynamical variation of the curvature of bending waveguides [21] or two-frequency modulations of refractive index [22] open additional opportunities for control of light propagation were appreciated.

   In this Letter we predict that by stacking waveguide arrays from segments with different frequencies of longitudinal refractive index modulation one can realize rich variety of switching scenarios. In particular, our results are useful for applications where light beam should occupy certain domains of the guiding structure over strictly required length and then should be displaced in a controllable way to another location inside the structure.

We describe the propagation of light beam along the $\xi$-axis of segmented waveguiding structure with the nonlinear Schrödinger equation for dimensionless light field amplitude $q$:

$$i\frac{\partial q}{\partial \xi} = -\frac{1}{2}\frac{\partial^2 q}{\partial \eta^2} - pR(\eta,\xi)q + \sigma q|q|^2, \qquad (1)$$

where $\eta, \xi$ are the normalized transverse and longitudinal coordinates, respectively; $p$ is the depth of refractive index modulation; the focusing nonlinearity is considered; while the function $R(\eta,\xi) = \sum_m [1+(-1)^m \mu \sin[\Omega(\xi)\xi]]Q(\eta-md)$ describes the profile of guiding structure composed out of $N$ identical waveguides with profiles $Q(\eta) = \exp(-\eta^6/a^6)$ and separation $d$; the parameter $\sigma = -1$ corresponds to a focusing nonlinearity, while $\sigma = +1$ corresponds to defocusing one. The refractive index in the adjacent waveguides is modulated along $\xi$ out-of-phase, with the frequency $\Omega(\xi)$ and the depth $\mu$. Here we suppose that the frequency $\Omega(\xi)$ changes in a step-like fashion, for example $\Omega = \Omega_1$ for $0 \leq \xi < \xi_1$ and for $\xi > \xi_2$, and $\Omega = \Omega_2$ for $\xi_1 \leq \xi \leq \xi_2$ (this is equivalent to embedding into modulated array of a single segment of the length $\delta\xi = \xi_2 - \xi_1$ with distinct modulation frequency). Analogously, one can add several segments with various lengths and frequencies into such arrays. Other laws of longitudinal frequency modulation may be considered, including linear variation from $\Omega_1$ to $\Omega_2$ values, but main results remain qualitatively similar. Further we set $a = 0.3$ (the waveguides with the width $\sim 3\ \mu\text{m}$), $d = 1.8$ (separation $\sim 18\ \mu\text{m}$), $\mu = 0.2$, and lattice depth $p = 11.6$ (that is equivalent to real refractive index contrast $\delta n \sim 10^{-3}$ at wavelength $\lambda = 800$ nm). We will be interested in the impact of parameters $\delta\xi$, $\Omega_{1,2}$ on propagation dynamics. Importantly, our continuous model, based on the nonlinear Schrödinger equation that describes evolution of linear and nonlinear excitations in multi-well dynamically modulated potentials, is also highly relevant for analysis of behavior of matter waves in Bose-Einstein condensates (in this case the evolution variable in Eq. (1) should be time rather than propagation distance) and it might be also illuminating in some problems of molecular dynamics. Notice also, that the control optical fields may adjust linear and nonlinear propagation regimes for a probe signal in the array of active waveguides. In particular, such fields may lead to periodic longitudinal variations of both linear refractive index and nonlinearity [23].

It is well established that single-frequency out-of-phase longitudinal refractive index modulation inhibits coupling between guiding channels even in linear regime [10]. Inhibition of light tunneling or coupling even for a simplest system of two channels is a resonant effect that occurs for the sequence of resonant modulation frequencies $\Omega_{rk}$. The highest frequency $\Omega_{r1}$ corresponds to the primary resonance, while secondary resonances are reached at the fractional frequencies that decrease approximately as $\Omega_{r1}/k$ with $k$. The resonant frequencies grow almost linearly with the depth of longitudinal modulation $\mu$. Further we consider segmented structure where $\Omega_1 = \Omega_{r1}$. This means that the laser beam entering one of the guides of such structure remains confined entirely in this channel as long as it propagates inside a resonant segment. However, when the beam enters the frequency-detuned (or non-resonant) segment where $\Omega_2 \neq \Omega_{rk}$ and coupling between waveguides becomes considerable, the input energy is redistributed between several channels. The final field distribution at the exit of nonresonant segment depends on its length, $\delta\xi$ and $\Omega_2$ parameters. When the resulting field enters the next resonant segment of the structure, it remains frozen since the coupling is inhibited again. In this way, one can achieve splitting of simple inputs into different configurations maintaining their internal structures with $\xi$ after leaving nonresonant segments - an effect that may be useful for optical communication systems. Notice that "continuous" model [Eq.(1)] that we use here takes into account a subtle dynamics of guided modes, which is completely ignored within the frame of simplified tight-binding approximation, which is frequently used upon analysis of dynamics of modulated systems [9]. In particular, the continuous model accounts for the important fact that the instantaneous modification of the frequency of longitudinal refractive modulation does not mean immediate modification of the coupling strength between adjacent waveguides. For instance, the step-like modification of modulation frequency results in excitation of slowly decaying oscillations of the mode width, with a frequency that generally does not coincide with new frequency of refractive index modulation. Such oscillations do affect switching dynamics. Therefore, real dynamics in modulated waveguide arrays is more complicated than simple tight-binding picture predicting that the dynamics of switching is governed by the effective coupling constant modified by zero-order Bessel function, whose roots are dictated by the frequency and amplitude of longitudinal refractive index modulation. Even this simple tight-binding model combined with Floquet-Bloch formalism leads to conclusion, that although the band of eigenvalues shrinks around the points corresponding to zeros of associated zero-order Bessel function (this shrinkage is an indication of tunneling inhibition), the width of this band remains finite, i.e. light tunneling inhibition in modulated arrays can be strong, but it cannot be complete [9]. In reality the degree of tunneling inhibition depends on numerous parameters, such as the depth of longitudinal refractive index modulation, separation, and widths of waveguides.

    Figure 1 illustrates this idea on the simplest example of coupler with single nonresonant segment, whose input and output facets are indicated by dashed lines. In all cases we used as an input beam $q|_{\xi=0} = Aw(\eta)$, where $A$ is the peak amplitude and $w(\eta)$ is the profile of linear mode of isolated waveguide. For parameters of our structure one has $\Omega_{r1} = 24.50\Omega_b$ and $\Omega_{r2} = 10.42\Omega_b$, where $\Omega_b = 2\pi/T_b$ is the frequency ($T_b$ is the period) of energy beating in unmodulated coupler. In Fig. 1(a) we set $\Omega_2 = \Omega_{r2}$, i.e. the modulation frequency jumps from the primary resonance to the secondary one. In this case, apparently, the light remains entirely localized in the input channel despite linear regime and only the frequency of small oscillations of beam amplitude diminishes inside the inserted segment. In Fig. 1(b) the coupling between guides results in equal distribution of energy flow between output channels because of small detuning of $\Omega_2$ from the secondary resonance. Further growth of modulation frequency $\Omega_2$ results in complete switching [Fig. 1(c)], followed by another event of equal energy splitting [Fig. 1(d)], and total power restoration in the original excited guide [Fig. 1(e)]. Thus, the setting considered here affords not only light localization in a desired channel [like in usual coupling inhibition scheme in Fig. 1(f)], but also enables control of the distance $\xi$ within which the light will be stored in a desired channel. Notice that shallow longitudinal refractive index modulations can be easily implemented in

practice in laser-written waveguide arrays [10]. Moreover, by using electro-optical effect and longitudinally segmented electrodes, etched on the top of straight waveguide array, or by means of properly modulated background illumination in photorefractive waveguide array, one can, in principle, flexibly control light switching or shearing scenarios in real time. The control of coupling between waveguides due to modification of frequency of shallow longitudinal refractive index modulations is apparently advantageous from technological point of view taking into account that in static waveguide arrays one has to dramatically change either refractive index or separation between waveguides in order to achieve the same modification of coupling strength.

We calculated the output energy sharing $S_{1,2} = U_{1,2}/U_{\text{in}}$ (here $U_{\text{in}} = \int_{-\infty}^{\infty} |q|^2 \, d\eta$ is the total input energy flow, and $U_{1,2}$ are output energy flows trapped in first and second channels) between channels of segmented coupler [Figs. 2(a) and 3(a)] and between first excited, second, and third channels (the definitions of $S_{1,2,3}$ are similar in this case) of truly infinite waveguide array [Fig. 2(c)] as a function of modulation frequency $\Omega_2$ for small input amplitudes $A \ll 1$. One can see that in the case of coupler $S_{1,2}$ are oscillating functions of $\Omega_2$, since for a given $\delta\xi$ value the number of switching events inside nonresonant segment may be different because the rate of light tunneling depends dramatically on $\Omega_2$. Increasing the length of nonresonant segment results in more complicated $S_{1,2}(\Omega_2)$ dependencies [compare Fig. 3(a) and 2(a) that were obtained for $\delta\xi = 1.1T_b$ and $\delta\xi = 2.4T_b$, respectively]. The dependencies $S_{1,2,3}(\Omega_2)$ obtained for infinite array are different from the oscillating curves obtained for coupler [Fig. 2(c)], because in infinite array the energy usually spreads over multiple waveguides. There exist different optimal $\Omega_2$ values corresponding to maximal concentration of energy flow in each of selected guides. As in the case of coupler, the complex field distribution that leaves nonresonant segment of array remains frozen in resonant part despite the fact that multiple waveguides are excited. Notice that in both cases of coupler and array $S_1$ attains maxima for $\Omega_2 = \Omega_{r1}, \Omega_{r2}$, when almost all energy remains in the input channel over entire propagation distance. The dependencies of energy sharing on the length of nonresonant segment are strictly periodic in the case of coupler [Fig. 2(b)] and appear as decaying oscillations in infinite waveguide array. If the detuning of $\Omega_2$ from either $\Omega_{r1}$ or $\Omega_{r2}$ increases (that is accompanied by growth of coupling strength) the period of oscillations decreases [Fig. 2(b)]. Notice that dependencies $S_k(\Omega_2)$, $k = 1,...,N$ for excitations in systems containing finite $(N > 2)$ number of waveguides may be really complex, with multiple maxima at specific $\Omega_2$ values corresponding to light localization (in some cases - almost complete) in different output channels (see Fig. 4). This indicates on new opportunities for control of output light distributions afforded by segmented structures.

By increasing the amplitude of input beam $A$, we also studied an impact of nonlinearity on the light field propagation in segmented guiding structures. In particular, in the case of coupler with focusing nonlinearity $(\sigma = -1)$, an increase of the input amplitude prevents the light from tunneling into second channel [compare the dependencies $S_{1,2}(\Omega_2)$ at $A = 0.01$ and $A = 0.27$ depicted in Figs. 3(a) and 3(b), respectively]. As a result, the maximal power fraction $S_2^{\max} = \max(S_2)$ that is trapped in the second channel rapidly diminishes with increase of input peak intensity $A^2$ [Fig. 3(c)]. Notice, that while similar nonlinear localization effect is possible in unmodulated structures, much smaller input intensities are required for single-guide trapping in longitudinally modulated systems. Interestingly, we found that defocusing nonlinearity ($\sigma = +1$) has very similar effect on the switching characteristics in our system. In somewhat counterintuitive fashion in defocusing medium an increase of peak amplitude also results in suppression of coupling into second channel, at least for $A < 0.4$. For fixed $A$ the dependencies $S_{1,2}(\Omega_2)$ obtained at $\sigma = \pm 1$ differ only by small-scale oscillations, which are out-of-phase for focusing and defocusing nonlinearities [Fig. 3(d)]. As a result the dependencies $S_2^{\max}(A^2)$ are almost identical for focusing and defocusing media.

Finally, the unusual switching scenarios that may be realized in segmented longitudinally modulated structures are presented in Fig. 4. In particular, in structures composed of finite number of waveguides (for example - four) a proper selection of the modulation frequency $\Omega_2$ and the nonresonant segment length $\delta\xi$ results in concentration and subsequent propagation of light practically in only one of the output channels which does not coincide with the input one [see Figs. 4(a) and 4(b)]. The rich variety of nontrivial switching scenarios is available in structures involving several alternating resonant and nonresonant segments. In particular, in Fig. 4(c) the light beam launched into the left channel of modulated coupler switches to other channel inside first nonresonant segment, stays in right channel over the length of subsequent resonant segment, and then splits into two identical beams on the second nonresonant segment.

Summarizing, we analyzed propagation of light in segmented longitudinally modulated waveguide arrays. In such structures composed of resonant and nonresonant segments one can achieve splitting of simple input beams into almost arbitrary patterns maintaining their internal structures with propagation after leaving nonresonant domains. Tuning the modulation frequencies and lengths of nonresonant segments in multichannel systems enables full control over redistribution and concentration of light in different output channels. It is worth noticing that all effects described here may be also observable in Bose-Einstein condensates.

# References with titles

# Figure captions

Figure 1. Propagation dynamics in two-channel coupler with step-like modification of modulation frequency at $A=0.01$, $\delta\xi=2.4T_b$. Dashed lines indicate the borders of segments with different modulation frequencies. The modulation frequency in central segment is (a) $\Omega_2=10.42\Omega_b$, (b) $11.01\Omega_b$, (c) $11.75\Omega_b$, (d) $12.70\Omega_b$, (e) $15.05\Omega_b$, and (f) $24.50\Omega_b$. In all cases $\Omega_1=24.50\Omega_b$.

Figure 2. Output energy sharing between channels of modulated coupler (a) versus $\Omega_2/\Omega_b$ at $\delta\xi=2.4T_b$ and (b) versus $\delta\xi/T_b$ at $\Omega_2=21.2\Omega_b$. (c) Output energy sharing for three neighboring channels in modulated waveguide array versus $\Omega_2/\Omega_b$ at $\delta\xi=2.0T_b$. In all cases $\Omega_1=24.5\Omega_b$ and $A=0.01$.

Figure 3. Output energy sharing between two channels of modulated coupler with focusing nonlinearity versus $\Omega_2/\Omega_b$ at $A=0.01$ (a) and $A=0.27$ (b). (c) Maximal output fraction of energy concentrated in second channel of coupler with focusing nonlinearity versus $A^2$. (d) The comparison of dependencies $S_{1,2}(\Omega_2)$ obtained at $A=0.25$ in focusing (black curves) and defocusing (red curves) media. In all cases $\delta\xi=1.1T_b$ and $\Omega_1=24.5\Omega_b$.

Figure 4. Propagation dynamics in a system of four channels containing single segment with nonresonant modulation frequency at $\delta\xi=2.0T_b$, $\Omega_2=15.30\Omega_b$. In (a) first channel was excited, while in (b) second channel was excited at $\xi=0$. (c) Propagation dynamics in coupler with two nonresonant segments of length $2.4T_b$ and $1.2T_b$ at $\Omega_2=11.75\Omega_b$. In all cases $A=0.01$ and $\Omega_1=24.50\Omega_b$.

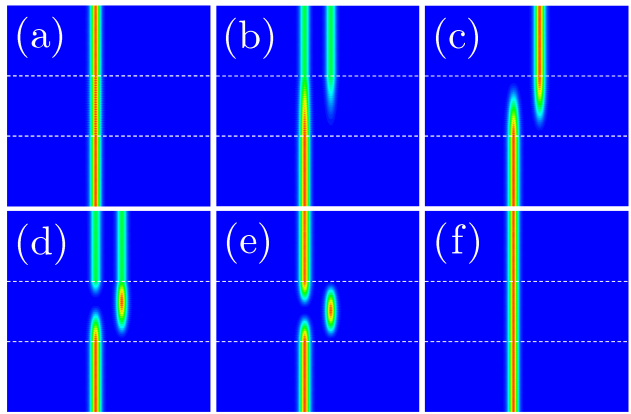

Figure 1. Propagation dynamics in two-channel coupler with step-like modification of modulation frequency at $A = 0.01$, $\delta\xi = 2.4 T_{\rm b}$. Dashed lines indicate the borders of segments with different modulation frequencies. The modulation frequency in central segment is (a) $\Omega_2 = 10.42\Omega_{\rm b}$, (b) $11.01\Omega_{\rm b}$, (c) $11.75\Omega_{\rm b}$, (d) $12.70\Omega_{\rm b}$, (e) $15.05\Omega_{\rm b}$, and (f) $24.50\Omega_{\rm b}$. In all cases $\Omega_1 = 24.50\Omega_{\rm b}$.

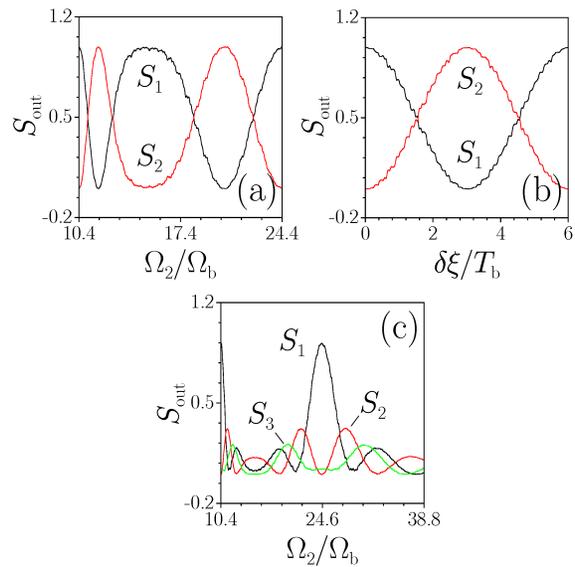

Figure 2. Output energy sharing between channels of modulated coupler (a) versus $\Omega_2/\Omega_b$ at $\delta\xi = 2.4T_b$ and (b) versus $\delta\xi/T_b$ at $\Omega_2 = 21.2\Omega_b$. (c) Output energy sharing for three neighboring channels in modulated waveguide array versus $\Omega_2/\Omega_b$ at $\delta\xi = 2.0T_b$. In all cases $\Omega_1 = 24.5\Omega_b$ and $A = 0.01$.

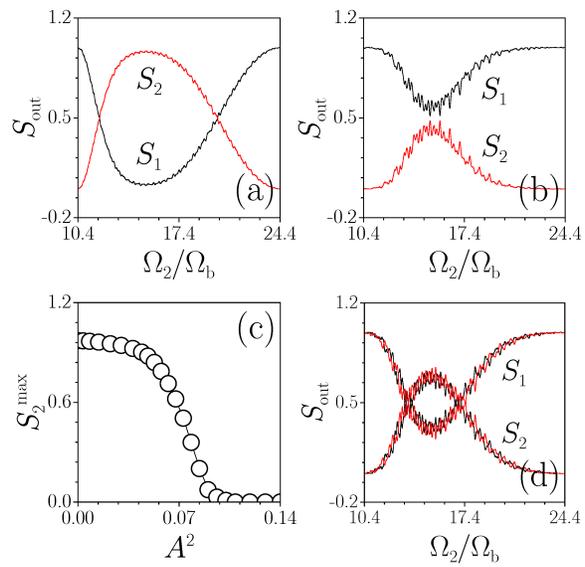

Figure 3. Output energy sharing between two channels of modulated coupler with focusing nonlinearity versus $\Omega_2/\Omega_b$ at $A=0.01$ (a) and $A=0.27$ (b). (c) Maximal output fraction of energy concentrated in second channel of coupler with focusing nonlinearity versus $A^2$. (d) The comparison of dependencies $S_{1,2}(\Omega_2)$ obtained at $A=0.25$ in focusing (black curves) and defocusing (red curves) media. In all cases $\delta\xi=1.1T_b$ and $\Omega_1=24.5\Omega_b$.

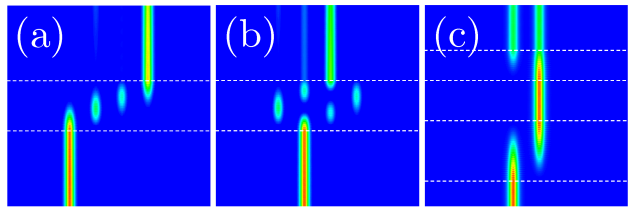

Figure 4. Propagation dynamics in a system of four channels containing single segment with nonresonant modulation frequency at $\delta\xi = 2.0 T_b$, $\Omega_2 = 15.30\Omega_b$. In (a) first channel was excited, while in (b) second channel was excited at $\xi = 0$. (c) Propagation dynamics in coupler with two nonresonant segments of lengths $2.4 T_b$ and $1.2 T_b$ at $\Omega_2 = 11.75\Omega_b$. In all cases $A = 0.01$ and $\Omega_1 = 24.50\Omega_b$.